\documentclass[twocolumn]{aastex631}

\usepackage{graphics,epsf}
\usepackage[utf8]{inputenc}
\usepackage{amsmath}                
\usepackage{amsfonts}               
\usepackage{amssymb}                
\usepackage{epsfig}                 
\usepackage{graphicx}
\usepackage{float}
\usepackage{color}

\usepackage[colorinlistoftodos]{todonotes}

\usepackage[para,online,flushleft]{threeparttable}

\newcommand{\cm}{{~\rm cm}}
\newcommand{\km}{{~\rm km}}
\newcommand{\s}{{~\rm s}}

\newcommand{\g}{{~\rm g}}
\newcommand{\G}{{~\rm G}}
\newcommand{\K}{{~\rm K}}
\newcommand{\erg}{{~\rm erg}}

\begin{document}

\title{Remnant masses of core collapse supernovae in the jittering jets explosion mechanism}
\date{December 2021}

\author[0000-0002-9444-9460]{Dmitry Shishkin}
\affiliation{Department of Physics, Technion, Haifa, 3200003, Israel; s.dmitry@campus.technion.ac.il; soker@physics.technion.ac.il}

\author[0000-0003-0375-8987]{Noam Soker}
\affiliation{Department of Physics, Technion, Haifa, 3200003, Israel; s.dmitry@campus.technion.ac.il; soker@physics.technion.ac.il}

\begin{abstract}
We conduct one dimensional (1D) stellar evolution simulations of non-rotating stars with initial masses in the range of $11-48 M_{\odot}$ to the time of core collapse and, using a criterion on the specific angular momentum fluctuations in the inner convective zones, estimate the masses of the neutron star (NS) remnants according to the jittering jets explosion mechanism. From the 1D simulations we find that several convective zones with specific angular momentum fluctuations of $j_{\rm{conv}} \ga 2.5 \times 10^{15} \cm^2 \s^{-1}$ develop near the edge of the iron core in all models. For this condition for explosion we find the NS remnant masses to be in the range of $1.3 -1.8 M_\odot$, while if we require twice as large values, i.e., $j_{\rm{conv}} \ga 5 \times 10^{15} \cm^2 \s^{-1}$, we find the NS remnant masses to be in the range of $1.4 -2.8 M_\odot$ (the upper values here might form black holes).
Note that in general the formation of black holes in the jittering jets explosion mechanism requires a rapidly rotating pre-collapse core, while we simulate non-rotating stars. 
\end{abstract}

\keywords{stars: jets – stars: massive – supernovae: general - stars: neutron}

\section{INTRODUCTION}
\label{sec:intro}

Stars with zero age main sequence mass of $M_{\rm ZAMS} \ga 8 M_{\odot}$ explode as core collapse supernovae (CCSNe) where the core collapses to a neutron star (NS) or a black hole (BH) as it releases huge amounts of gravitational energy (e.g., \citealt{Janka2012}).
The collapsing core material encounters a shock, the stalled shock, at a radius of $\simeq 100 \km$ before it settles onto the newly born NS.  
There are two theoretical explosion mechanisms to utilize some small fraction of this gravitational energy to explode the star, the delayed neutrino explosion mechanism \citep{BetheWilson1985} and the jittering jets explosion mechanism \citep{Soker2010}.
Both of these mechanisms require perturbations in the pre-collapsing core to facilitate the explosion.
The delayed neutrino explosion mechanism requires perturbations to break the spherical symmetry behind the stalled shock, which facilitate the revival of the stalled shock (e.g., \citealt{CouchOtt2013, OConnorCouch2018, Mulleretal2019Jittering, Couchetal2020, Burrowsetal2020, KazeroniAbdikamalov2020, Burrows2021review, Vartanyan2022}).
  
In the jittering jets explosion mechanism, which we adopt in this paper, these perturbations serve as the source of stochastic angular momentum fluctuations (\citealt{Soker2019SASI, Soker2019JitSim}) that are then further amplified by instabilities. The main relevant instability is the spiral standing accretion shock instability (spiral SASI, e.g., \citealt{BlondinMezzacappa2007, Iwakamietal2014, Kurodaetal2014, Fernandez2015, Kazeronietal2017} for studies of the SASI, and, e.g.,  \citealt{Andresenetal2019, Walketal2020, Nagakuraetal2021, Shibagakietal2021},  for recent simulations that demonstrate the spiral SASI).
According to the jittering jets explosion mechanism the amplified perturbations lead to the formation of intermittent accretion disks (or belts, \citealt{SchreierSoker2016}), which in turn launch the jittering jets (e.g., \citealt{PapishSoker2011,  GilkisSoker2014, GilkisSoker2015, Quataertetal2019}).
Supplying large enough perturbation seeds to the spiral SASI such that the final angular momentum fluctuations form the intermittent accretion disks is the key challenge of the jittering jets explosion mechanism.

When accretion is from the hydrogen-rich envelope (e.g., \citealt{Quataertetal2019, AntoniQuataert2021}) or from the helium-rich shell above the core (e.g., \citealt{GilkisSoker2014}) that are at large distances from the center, the angular momentum fluctuations are sufficient by themselves to form intermittent accretion disks. 
In these cases the large masses inner to these shells imply the formation of a BH remnant. 
We here concentrate on explosions that leave a NS remnant (or a BH just above the maximum allowed mass for NSs). 

In both explosion mechanisms the seeds of these perturbations in slowly-rotating cores (or not rotating at all) come from the convective zones of the pre-collapsing core. Although the two explosion mechanisms are different, i.e., by their prediction of which stars explode and on the possible explosion energies (e.g., \citealt{GofmanSoker2020}), both processes might work at exploding CCSNe. \cite{Soker2019JitSim} argued, based on some simulations of the neutrino delayed explosion mechanism (e.g., \citealt{Mulleretal2017, Mulleretal2018Bipol, Mulleretal2019Jittering}), that most likely there is a mutual influence between neutrino heating and accreting matter with stochastic angular momentum, and that both operate together to bring the cores of CCSNe to explode by variable bipolar outflows, i.e., jittering jets.  In a recent study \cite{Soker2022Boosting} further develops this synergy between jittering jets and neutrino heating and estimates that neutrino heating doubles the energy of the outflow that the jets induce, and thus plays a role in the jittering jets explosion mechanism in boosting the energy of the outflow that the jets trigger. We also note that from a radius of about $3000 \km$ the jittering jets drive a shock that expands to the oxygen and outer core layers in a similar manner to that of the delayed neutrino mechanism, although the symmetry might be different (e.g., \citealt{Soker2018}). Such a shock will lead to  nucleosynthesis in the jittering jets explosion mechanism that is similar to that of the delayed neutrino mechanism (\citealt{Soker2018}). 

In the jittering jets explosion mechanism the launching of jets starts when the inner convective zone is accreted by the proto-NS. This takes place at $\simeq 0.1 - {\rm few} \times 0.1 \s$ after the stalled shock establishes itself at a radius of about 100~km. The duration of each jet-launching episode is $\simeq 0.01-0.1 \s$, and the energy the jets carry in each such episode is $\approx {\rm few} \times 10^{49} - {\rm few} \times 10^{50} \erg$ (e.g., \citealt{PapishSoker2011}). The jet-activity phase starts at about the same time that the neutrino driven mechanism revives the stalled shock in successful explosions (e.g., \citealt{Burrowsetal2020}). However, jets activity might continue for a longer time in cases where accretion continues even after the revival of the stalled shock, e.g., from the equatorial plane in case of a rotating core.

The jittering jets explosion mechanisms has several advantages. It might account for the morphology of some supernova remnants that have the signatures of jets (e.g., \citealt{GrichenerSoker2017, Soker2022SNR}). Also, it connects to the explosion of very energetic CCSNe that most probably require jets for the explosion via the jet feedback mechanism \citep{Soker2016Rev}. The jittering jets explosion mechanism predicts that there are no failed CCSNe, even in the formation of a BH. This is compatible with the recent results of \cite{ByrneFraser2022}, who in a systematic search find no transients that are consistent with failed CCSNe.  
However, there are no simulations yet that present the jittering jets explosion mechanism. Such simulations must be of very high resolution and include magnetic fields \citep{Soker2019SASI}.

There are several three dimensional (3D) simulations of the convection in the inner zones of pre-collapsing cores (e.g., \citealt{FieldsCouch2020, FieldsCouch2021, Yoshida2021}). \cite{FieldsCouch2020} compared their 3D simulations with their 1D simulations. Their results show that in the 3D simulations the amplitudes of the velocity convective fluctuations are about 2-4 times larger than those in the 1D simulations. \citealt{FieldsCouch2021} and \cite{Yoshida2021} also find that large scale modes (low values of spherical harmonic orders $l$) are prevalent in the inner convective zones.

In our previous study \citep{ShishkinSoker2021} we followed the evolution of stars to core collapse velocities of $v_{\rm fall} > 1000 \km \s^{-1}$. We argued, when we consider the larger expected velocity variations in the 3D simulations of \cite{FieldsCouch2020}, that the convective specific angular momentum fluctuations in the core are sufficiently large seed-perturbations to SASI for the formation of intermittent accretion disks around the newly born NS. We adopt this claim in the present study. 

In the present study we use 1D stellar evolutionary code (section \ref{sec:CoreCollapse}) to connect the properties of the pre-collapse core with the expected final remnant mass in the frame of the jittering jets explosion mechanism (section \ref{sec:Results}). There are studies that make this connection in the frame of the neutrino driven explosion (e.g., \citealt{Pattonetal2021}). Our results differ from theirs, e.g., in the jittering jets explosion mechanism there are no failed CCSNe and single stars do not form massive BHs, as we discuss in section \ref{sec:summary}.

\section{The numerical scheme}
\label{sec:CoreCollapse} 

We use the stellar evolution code \textsc{mesa} (version 10398; \citealt{Paxtonetal2010, Paxtonetal2013, Paxtonetal2015, Paxtonetal2018, Paxtonetal2019}) to follow stars to core collapse.
We examine a spectrum of initial masses in the range of $M_{\rm ZAMS}= 11 - 48 M_\odot$ and with metalicity of $z = 0.02$. We do not include rotation.

We use the numerical prescription \textit{(inlist)} as in our earlier study \cite{ShishkinSoker2021}, which is similar to the \textit{inlist} of \cite{FieldsCouch2020} who compared 1D with 3D simulations of a collapsing $M_{\rm ZAMS}= 15 M_\odot$ star.

Since we focus our study on the last several seconds of collapse, we terminate our simulations when the maximum infall velocity reaches a value of $v_{\rm fall,m}>4000 \km \s^{-1}$. The region that first reaches this velocity is the outer edge of the iron core. 

Below we explain the numerical details and elaborate on some of the numerical issues and difficulties related to the fast collapse.

\textit{Convective velocity.} The numerical code \textsc{mesa} calculates the convective velocity by the mixing length theory. 
We set $\alpha_{ML}=1.5$.
This simplified theory does not catch the complicated convective motion in reality. For example, \cite{FieldsCouch2021} find the maximum convective velocity in their three-dimensional simulations to be about two to four times larger than the velocity that \textsc{mesa} gives. One possible outcome is that the exact convective velocities and convective zones are sensitive to the numerical prescription. Two stellar models that differ only slightly by their mass, might have different convective zones and convective velocities. 

\textit{Nuclear Network.} For the nuclear reaction network we use the `approx21' net (a 21 isotope network which is standard for massive stars) as a compromise between simulating resources and accuracy. \cite{Farmer2016} performed an in-depth analysis of the effects of different networks on the outcomes of massive star evolution, from which we conclude that with sufficiently fine mesh refinement the 21 isotope network is adequate. {{{{ However, a larger nuclear network, e.g., 127 isotopes as \cite{Farmer2016} use, will yield somewhat more realistic results. }}}}

\textit{Temporal Resolution.} We change several numerical parameters that control the size of a time-step from their default values. In many cases we had to manually intervene with the simulation time-steps because of numerical difficulties. To overcome some convergence difficulties we separate each simulation to two parts, one until the iron core reaches a mass of $M_{\rm core}^{\rm Fe} \simeq 1.2M_\odot$, and the other after that time. In the second part we use stronger constraints on the time-steps. For example, we change the value of $varcontrol\_target=1d-4$ in the first part to $varcontrol\_target=1d-5$ in the second part. Changing the length of the time-steps lead to somewhat different convective zones and velocities, but with the very short time-steps of our simulations we reached a consistent behavior. 

\textit{Mesh Refinement.} We set our one-dimensional maximum cell mass to be $10^{-4} M_{\rm{star}}$ ($max\_dq<1d-4$), where $M_{\rm star}$ is the stellar mass. This is two orders of magnitude higher resolution than the default of \textsc{mesa}. 
We found that at this maximum shell mass we reach more or less numerical convergence, i.e., smaller numerical calls do not change much the results in the relevant convective zones (see Appendix \ref{appendix1}). The maximum number of cells we have used in most simulations is around $2 \times 10^4$ (in appendix \ref{appendix1} we compare to simulations with higher resolutions).

Encouraging to our study is that although some convective zones disappear and reappear in the last tens of seconds of the collapse, the inner convective zone, which is the most relevant to us is not sensitive to the resolution (see Appendix \ref{appendix1}).  

\textit{Other Parameters:} We set the wind mass loss rate to follow the \rm{Dutch} scheme in \textsc{mesa} with a wind scaling parameter of $0.8$. 

We set the overshoot parameters to $f=0.004$ and $f_0=0.001$ for most runs. In four runs, for $M_{\rm ZAMS}=11,11.5, 12$ and $39 M_\odot$, we set $f=0.01$ and $f_0=0.002$ because of numerical convergence difficulties. In these runs we also modified the wind scaling factor to $1$, which results in lower envelope masses (see Appendix \ref{appendix2}). These different numerical parameters did not have a substantial effect on the convective profile as we learn from comparing to stellar models with adjacent masses.

We turn on the Ledoux criteria with its usual values of $\rm{alpha\_semiconvection} = 0.01$ and $\rm{thermohaline\_coeff} = 1$.

Because we monitor the collapse we also enable hydrodynamical radial velocity.

We use the following criteria to include convective zones. First, in each convective zone we calculate the specific angular momentum fluctuations $j_{\rm conv}(m)=v_{\rm conv}(m)r$ and average the convective velocity $v_{\rm conv}(m)$ over the numerical cells (numerical shells) in the convective zone.  
We include only convective zones that obey all the following four conditions. (1) An average convective velocity of $\bar{v}_{\rm conv} > 3 \km \sec^{-1}$. (2) A  convective zone mass of $ \Delta m_{\rm conv} \ge 0.01$. (3) The convective zone exists at least at one time within 20 seconds from collapse, i.e., during $t_3 > -20 \s$. (4) Either the convective zone exists at the moment we terminate the simulation, or, if the convective zone disappears just before we terminate the simulation, the convective velocity, at least at one time during the time period of $t_3 > -20 \s$, satisfies 
\begin{equation}
v_{\rm fall} (m) > v_{\rm conv} (m),  
\label{eq:Cond4}
\end{equation}
i.e., infall velocity in the convective zone is larger than the convective velocity. The latter condition implies that the convective motion does not have time to decay until the zone reaches the center. 
 
\section{Results}
\label{sec:Results}

\subsection{Angular momentum criterion}
\label{subsec:Res}

Like in \cite{ShishkinSoker2021} we use the value of the specific angular momentum fluctuations $j_{\rm conv}(m) = v_{\rm conv}(m) r$ to locate the mass shell at which explosion takes place. We take this to represent the value of the specific angular momentum of the mass that the newly born NS accretes, but emphasise that there are two opposite effects that change the angular momentum of the accreted mass. Firstly, the stochastic convective motion implies that 3D averaging over different convective elements will give a lower values of the specific angular momentum of the accreted mass (e.g., \citealt{GilkisSoker2015, Quataertetal2019}). On the other hand, according to the recent version of the jittering jets explosion mechanism (e.g., \citealt{Soker2019SASI}) the spiral-SASI modes (see section \ref{sec:intro}) substantially amplify the initial perturbations of the convection motion.
In addition, the three dimensional hydrodynamical simulations of \cite{FieldsCouch2021} show that the peak of the convective velocity is about 2-4 times larger than what the mixing length theory gives, $j_{\rm 3d} (m) \simeq (2-4)j_{\rm conv} (m)$, and that the convective elements are large (implying smaller number of convective elements to average over).   
For these arguments we simply adopt our earlier study \citep{ShishkinSoker2021} and consider the condition for explosion in the frame of the jittering jets explosion mechanism (neglecting altogether any core rotation) to be (see more arguments in that paper) 
\begin{equation}
\bar{j}_{\rm conv,01} \ga j_{\rm jje} \simeq {\rm few} \times 10^{15} \cm^2 \s^{-1}, 
\label{eq:JJE}
\end{equation} 
where $j_{\rm jje}$ is the minimum value to set CCSN explosions in the frame of the jittering jets explosion mechanism and $\bar{j}_{\rm conv,01}$ is defined as follows. We examine the inner convective zone to find a spherical shell of mass $0.01 M_\odot$ within the convective zone that has the largest value of the average of ${j}_{\rm conv}(m)$ over a $0.01 M_\odot$ shell. 
   
To find the value of $j_{\rm conv} (m)$ we follow the collapsing core after the iron core reaches a mass of $M^{\rm Fe}_{\rm core} \simeq 1.2M_\odot$. At each time step we examine the value of the convective velocity and specific angular momentum as function of mass coordinate $v_{\rm conv}(m,t_3)$ and $j_{\rm conv} (m,t_3)$, where $t_3$ is the time relative to the evolutionary time when the maximum collapse velocity in the core reaches $v_{\rm fall,m}=1000 \km \s^{-1}$. From these we average $j_{\rm conv} (m,t_3)$ over each convective zone, and find the mass coordinates of the convective zone where $\bar j_{\rm conv}(t_3)$ exceeds the prescribed value $j_{\rm jje}$. We focus here on $ j_{\rm jje} = 2.5 \times 10^{15}$ and $j_{\rm jje} =5 \times 10^{15}$. We consider the perturbations in a convective zone to survive down to the stalled shock near the center if the infall velocity of that convective zone is larger than the convective velocity (equation \ref{eq:Cond4}). 

We take the baryonic remnant mass of the CCSN to be the mass inner to the smallest mass coordinate where our explosion condition in equation (\ref{eq:JJE}) holds at least at one time during the last several seconds of collapse 
\begin{equation}
M_{\rm rem,B} = m(\bar j_{\rm conv,01}>j_{\rm jje}).
\label{eq:Mremnant}
\end{equation}
We compute the final gravitational mass $M_{\rm rem,G}$ from the baryonic mass $M_{\rm rem,B}$ using equation 30 from \cite{Lattimer2001} with a neutron star radius of $R_{NS}=12 \km$ as in equation 9 in \cite{Sukhbold2016}.

A few words on the spiral-SASI are in place here. 
\cite{Burrowsetal2020} conduct 3D simulations of several CCSNe, from a lower mass of $9 M_\odot$ to an upper mass of $60 M_\odot$.
Their results show that in the models that explode the spiral-SASI did not develop. The spiral-SASI did develop in the three models that did not explode. We note the following. (1) \cite{Burrowsetal2020} introduce velocity perturbations with amplitudes of $100 \km \s^{-1}$. We expect the perturbations to be more than an order of magnitude larger. \cite{MullerJanka2015}, for example,  introduced perturbations that have amplitudes of $\simeq 2000 \km \s^{-1}$ and do get strong shear around the proto-NS. More recently, \cite{Bolligetal2021} find in their simulation of an exploding model that perturbations from the pre collapse oxygen burning shell add to the driving of large-scale, non-radial fluid motions (convective overturn or SASI) in the postshock layer. (2) Some simulations do get vigorous spiral-SASI (e.g., \citealt{Bolligetal2021}). A recent example is the appearance of spiral-SASI in a $70 M_\odot$ model before BH formation in the simulations by \cite{Shibagakietal2021}. Other studies do find vigorous spiral-SASI that increases with increasing pre-collapse core rotation (e.g., \citealt{Jankaetal2016}), even in moderate rotation as predicted by stellar evolution (e.g., \citealt{Blondinetal2017}). (3) The SASI itself appears also in the simulations by \cite{Matsumotoetal2022} that include magnetic fields, and disappear after shock revival.

The point from the discussion above, that there is no consensus on the appearance and magnitude of SASI. However, it seems that before explosion the SASI modes might exist in many cases. If neutrino heating does not derive the explosion then, we argue, jittering jets (with boosting from neutrino heating; \citealt{Soker2022Boosting}) will explode the star as the large pre-collapse convective perturbations with spiral-SASI lead to accretion disk/belt formation.

\subsection{Remnant masses}
\label{subsec:RemMass}

\begin{figure*}
\begin{center}
\includegraphics[trim=5cm 0cm 1cm 0cm,scale=0.45]{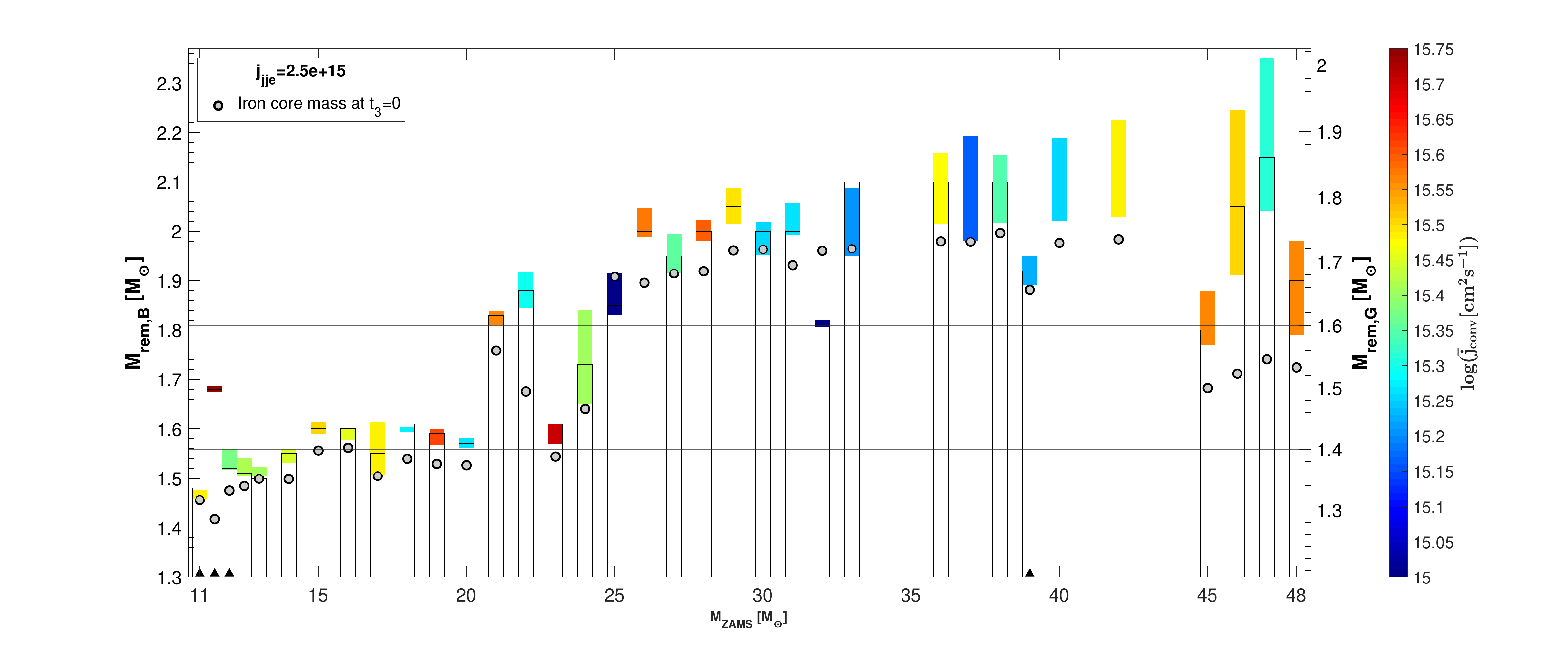}
\caption{The expected remnant masses in the frame of the jittering jets explosion mechanism for a minimal specific angular momentum fluctuations to cause explosion of $j_{\rm jje}= 2.5 \times 10^{15} \cm^2 \s^{-1}$. The horizontal axis is the initial stellar mass $M_{\rm ZAMS}$. The vertical axes are for the final remnant mass given by the height of the each column, the left axis for the final baryonic mass, i.e., the mass of the collapsing core, and the right axis for the final gravitational mass of the remnant. The colored region of each stellar model is the extent of the most inner convective zone that obey our conditions to seed explosion. The colors depict the average value of specific angular momentum fluctuations of the convective motion $\bar{j}_{\rm conv}$ according to the color-bar on the right. Note that the value $\bar j_{\rm conv}$ is lower than $\bar j_{\rm conv,01}$ that we use for the criterion in equation (\ref{eq:JJE}). The gray-filled black-circles show the iron core mass at  $t_3=0$, i.e., when the maximum collapsing velocity is $v_{\rm fall,m}=1000 \km \s^{-1}$. The horizontal lines at $M_{\rm rem, G}=1.4$, $1.6$ and $1.8 M_\odot$ serve to guide the eye and for comparison with Fig. \ref{fig:RemMass2}
}
\label{fig:RemMass}
\end{center}
\end{figure*}

\begin{figure*}
\begin{center}
\includegraphics[trim=5cm 0cm 1cm 0cm,scale=0.45]{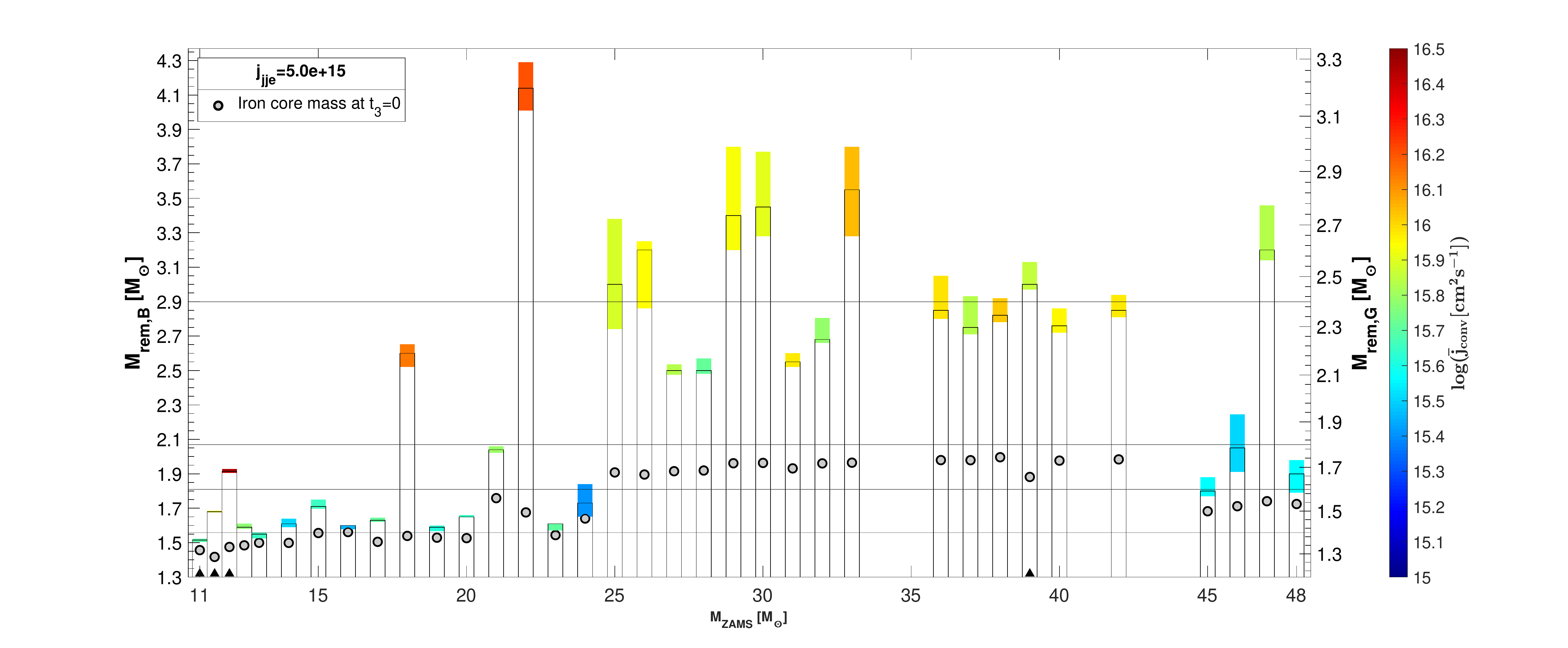}
\caption{Similar to  Fig \ref{fig:RemMass} but for $j_{\rm jje}=5 \times 10^{15} \cm^2 \s^{-1}$. Note the different scales of the vertical axes and of the color-bar of Fig. \ref{fig:RemMass} and of Fig. \ref{fig:RemMass2}. }
\label{fig:RemMass2}
\end{center}
\end{figure*}

In Figs \ref{fig:RemMass} and \ref{fig:RemMass2} we show our main results, which are our predicted remnant masses according to the jittering jets explosion mechanism (equation \ref{eq:Mremnant}) and for a minimal angular momentum fluctuations value of $j_{\rm jje}=2.5 \times 10^{15 }\cm^2 \s^{-1}$ and $j_{\rm jje}=5 \times 10^{15} \cm^2 \s^{-1}$, respectively.
On the left axis we display the expected baryonic remnant mass, whereas the right axis gives the gravitational remnant mass.
Note that the relation between $M_{\rm rem,B}$ and $M_{\rm rem,G}$ is not linear. In the figures the axis of $M_{\rm rem,B}$ is linear, but not that of $M_{\rm rem,G}$.
The colors indicate the value of $\bar{j}_{\rm conv}$ in the convective layer that has $\bar{j}_{\rm conv,01} > j_{\rm jje}$. Note again that $\bar{j}_{\rm conv}$ is the average of ${j}_{\rm conv}$ over the entire convective zone, while $\bar{j}_{\rm conv,01}$ is the maximum of an average over a shell of $0.01M_\odot$ inside that convective zone (equation \ref{eq:JJE}). 
Horizontal lines at $M_{\rm rem,G}=1.4,1.6,1.8,2.4$ serve as visual guides and for comparison between Figs \ref{fig:RemMass} and \ref{fig:RemMass2}.
We also mark by gray-filled black-circle the iron core mass at $t=t_3$ ($t_3=0$).
Note that the left vertical axes of $M_{\rm rem,B}$ starts from $1.3 M_\odot$ and that the colors and extend of the vertical axes are different in the two figures. The black triangles at the base of four columns indicates some numerical adjustments for these stellar models as we explain in section \ref{sec:CoreCollapse}.  

Before we list the main results, we note from Figs. \ref{fig:RemMass} and \ref{fig:RemMass2} that there are non-monotonic variations of the specific angular momentum fluctuations and the remnant masses between models close in $M_{\rm ZAMS}$. This results from the sensitivity of the properties of the collapsing core to some numerical parameters (more in appendix \ref{appendix1}). This demonstrates the uncertainties in the values of our results. Nonetheless, a clear picture emerges from our results, but one should keep these uncertainties in mind. 

The main properties of the remnant masses are as follows. By remnant mass we refer here to the gravitational mass which is the measured mass of NSs. 
\begin{enumerate}
\item \textit{Specific angular momentum.} We set the minimum value of the specific angular momentum fluctuations to be $j_{\rm jje}=2.5 \times 10^{15} \cm^2 \s^{-1}$ in Fig. \ref{fig:RemMass} and $j_{\rm jje}=5 \times 10^{15} \cm^2 \s^{-1}$ in Fig. \ref{fig:RemMass2}. However, we note in both figures that in most cases the actual value of $\bar{j}_{\rm conv,01}$ is even larger, making explosion more likely in the jittering jets explosion mechanism. 
\item \textit{Iron core mass.} In many cases, but not in all, the expected NS remnant mass is similar to the mass of the iron core. For minimum specific angular momentum fluctuations of $j_{\rm jje}=2.5 \times 10^{15} \cm^2 \s^{-1}$ (Fig. \ref{fig:RemMass}) this is the case for most models, while for $j_{\rm jje}=5 \times 10^{15} \cm^2 \s^{-1}$ (Fig. \ref{fig:RemMass2}) this is true mainly for $M_{\rm ZAMS} \la 24 M_\odot$. In the present study the similarity of the iron core mass and the NS remnant mass results from the strong convection at the edge of the iron core. 
\item \textit{Remnants of $M_{\rm rem,G} \simeq 1.3 M_\odot$.} The stellar model of $M_{\rm ZAMS} = 11 M_\odot$ yields a NS remnant of $M_{\rm ren,G} \simeq 1.3 M_\odot$. We will study the range of lower mass stars, many of which will suffer electron capture SNe, in a forthcoming paper.  
\item \textit{The remnants of $12 M_\odot \la M_{\rm ZAMS} \la 24 M_\odot$ stellar models.} Most stellar models with ZAMS mass of $12 M_\odot \la  M_{\rm ZAMS} \la 24 M_\odot$ leave a NS mass remnant of $M_{\rm rem,G} \simeq 1.4 M_\odot$.
\item \textit{The remnants of $25 M_\odot \la M_{\rm ZAMS} \la 50 M_\odot$ stellar models.} Most stellar models with ZAMS mass of $25 M_\odot \la M_{\rm ZAMS} \la 50 M_\odot$ leave a NS mass remnant of $M_{\rm rem,G} \simeq 1.6-1.8 M_\odot$ in the case of  $j_{\rm jje}=2.5 \times 10^{15} \cm^2 \s^{-1}$ (Fig. \ref{fig:RemMass}) and $M_{\rm rem,G} \simeq 2.0-2.8 M_\odot$ in the case of  $j_{\rm jje}=5 \times 10^{15} \cm^2 \s^{-1}$ (Fig. \ref{fig:RemMass2}). Likely, most form a NS and not a BH. In section \ref{sec:summary} we discuss the formation of BHs in the frame of the jittering jets explosion mechanism. 
\item \textit{Thin convective zones.} In some cases the convective zones are very thin. We give two examples from Fig. \ref{fig:RemMass}. In the case of $M_{\rm ZAMS}=11 M_\odot$ the inner convective zone (the one we show) at $m_1=1.47 M_\odot$ has a mass of $\Delta m_1=0.015 M_\odot$, and there is another close convective zone at $(m,\Delta m)_2=(1.515 M_\odot, 0.015 M_ \odot)$. In total the convective mass is $\Delta m_{\rm conv} = 0.03 M_\odot$. As another example, the mass in the convective zone of the $M_{\rm ZAMS} = 18 M_\odot$ is $\Delta m_{\rm conv} = 0.01 M_\odot$. There is another close convective zone but with a somewhat lower value of $j_{\rm conv} = 2 \times 10^{15} \cm^2 \s^{-1}$ (80\% of the minimum value we require) at $(m,\Delta m)_2=(1.66 M_\odot, 0.03 M_ \odot)$. We discuss the implications of these thin convective zones below. 
\end{enumerate}

A convective zone with a mass of only $\Delta m_{\rm conv}=0.03 M_\odot$ cannot lead to explosion by itself. For example, if the newly born NS launches a mass fraction of $f_{\rm jets}=0.1$ in the jets at a velocity of $v_{\rm j} = 150, 000 \km \s^{-1}$, the energy in the jets is $E_{\rm jets} = 7 \times 10^{50} \erg$. This is marginal for typical explosions. However, as we discussed in section \ref{subsec:Res}, the convective stochastic motion forms the perturbation seeds that instabilities, like the spiral-SASI (section \ref{sec:intro}), amplify in the regions inner to the stalled shock of the infalling gas. We expect the instabilities to both increase the amplitude of the specific angular momentum fluctuations and to be active for some time when gas from non-convective zones enters the instability zone.
To confirm our claims for the remnant masses, and more generally the jittering jets explosion mechanism, future high-resolutions simulations that include magnetic fields will have to demonstrate our conjecture/requirement for the amplifications of the specific angular momentum fluctuations by the spiral-SASI.
   
\section{Summary and discussion}
\label{sec:summary} 

In our \textsc{mesa} simulations of the $M_{\rm ZAMS} \simeq 11-48 M_\odot$ range with metalicity of $z=0.02$ we find that inner convective zones in the collapsing core, often at the edge of the iron core, with specific angular momentum fluctuations of the order of $\ga 2.5 \times 10^{15} \cm^2 \s^{-1}$ are consistently formed.
Scaling these to the results of 3D simulations (e.g., \cite{FieldsCouch2021}; section \ref{subsec:Res}) and considering that instabilities, such as the spiral-SASI (section \ref{sec:intro}), further amplify the seed perturbations, we claim that the specific angular momentum fluctuations that we find are sufficiently large to form stochastic intermittent accretion disks that launch jittering jets. 
These jittering jets explode the star according to the jittering jets explosion mechanism, or at least trigger the explosion that is then boosted with neutrino heating \citep{Soker2022Boosting}.

The inner convective zone that satisfies $\bar{j}_{\rm conv,01} > j_{\rm jje}$ (equation \ref{eq:JJE}) implies, under our assumptions, a remnant mass according to equation (\ref{eq:Mremnant}). 
For $j_{\rm jje}=2.5 \times 10^{15} \cm^2 \s^{-1}$ (Fig. \ref{fig:RemMass}) we find NS gravitational remnant masses of $M_{\rm rem,G} \simeq 1.3 M_\odot$ for the lower mass stellar model of $M_{\rm ZAMS}= 11 M_\odot$ and up to $M_{\rm rem,G} \simeq 1.8 M_\odot$ for the higher mass stellar models. For $j_{\rm jje}=5.0 \times 10^{15} \cm^2 \s^{-1}$ (Fig. \ref{fig:RemMass2}) the range is $M_{\rm rem,G} \simeq 1.4 M_\odot$ to $M_{\rm rem,G} \simeq 2.8 M_\odot$ with one exception.
There are uncertainties to these values as we discussed in section \ref{subsec:RemMass}. 
Overall, the remnant masses we find here according to the jittering jets explosion mechanism are consistent with observed masses of isolated NSs (\citealt{Meskhi2021}).

Our results apply only to single non-rotating stars. Pre-collapse core rotation can change the remnant mass in two ways. Consider a core rotation such that the specific angular momentum of the material at the edge of the iron core is $j_{\rm Fe,rot}$. If $j_{\rm Fe,rot}$ is not much larger than the convective angular momentum fluctuation in that region, namely $j_{\rm Fe,rot} \la {\rm few} \times j_{\rm conv}$, the jets that the newly born NS launches jitter. The jittering jets interact with the entire core and explode the star. The rotation can ease the condition for accretion disk formation, and might result in a somewhat lower mass remnant than what we find for single non-rotating stars. This might explain the finding that NSs in binary systems are on average lighter than isolated NSs (e.g., \citealt{Meskhi2021,Schwab2010}). 

If on the other hand $j_{\rm Fe,rot} \gg j_{\rm conv}$ and $j_{\rm Fe,rot} \ga 2 \times 10^{16}$ the jets do not jitter much. The first condition implies that the angular momentum fluctuations of the accreted mass do not change much the jets' axis, and the second condition implies that the specific angular momentum of the accreted mass is sufficient to form a persistent accretion disk around the newly born NS. \citep{Gilkisetal2016} argue that under these conditions the jets maintain a more or less constant axis (see also \citealt{Soker2017}), and although they eject all the core and envelope gas along the polar directions, they do not expel mass from near the equatorial plane. This is a process of inefficient jet feedback mechanism (JFM), implying that the core continues to accrete mass from the large volume of near the equatorial plane, and can grow to form a BH. As the jets continue to operate and expand along the polar directions, the total energy they carry can be tens to hundreds of times the binding energy of the core (an inefficient JFM). 

The points to take from the above discussion are as follows. ($i$) In the jittering jets explosion mechanism there are no failed SNe. To the contrary, the formation of BHs might result in the most energetic CCSNe \citep{Gilkisetal2016, Soker2017}. ($ii$) Even if, contrary to our claim here, the inner core does not manage to form the jittering jets to explode the star, the outer parts of the core and the inner parts of the convective envelope will definitely do so as the specific angular momentum fluctuations there are very large (e.g., \citealt{GilkisSoker2014, Quataertetal2019, AntoniQuataert2021}; see Fig. \ref{fig:App2fig} in appendix \ref{appendix2}). Therefore, the formation of a BH requires a rapid pre-collapse core rotation to ensure non-jittering jets. The large core-angular momentum requires either evolution with little mass loss or a binary interaction.  
 
The simulations of very massive stars of $M_{\rm ZAMS} \ga 50 M_\odot$, which require caution (e.g., \citealt{Agrawaletal2021}) and of electron capture SNe, as well as the inclusion of core rotation, are the subjects of future studies.

\section*{Acknowledgments}
We thank an anonymous referee for helpful comments. This research was supported by a grant from the Israel Science Foundation (769/20).

\section*{Data availability}
The data underlying this article will be shared on reasonable request to the corresponding author.  

\appendix

\setcounter{figure}{0}                  
\renewcommand\thefigure{A.\arabic{figure}} 

\section{Resolution study}
\label{appendix1}

\begin{figure*}
\begin{center}
 \includegraphics[trim=2cm 1cm 1cm 2cm,scale=0.4]{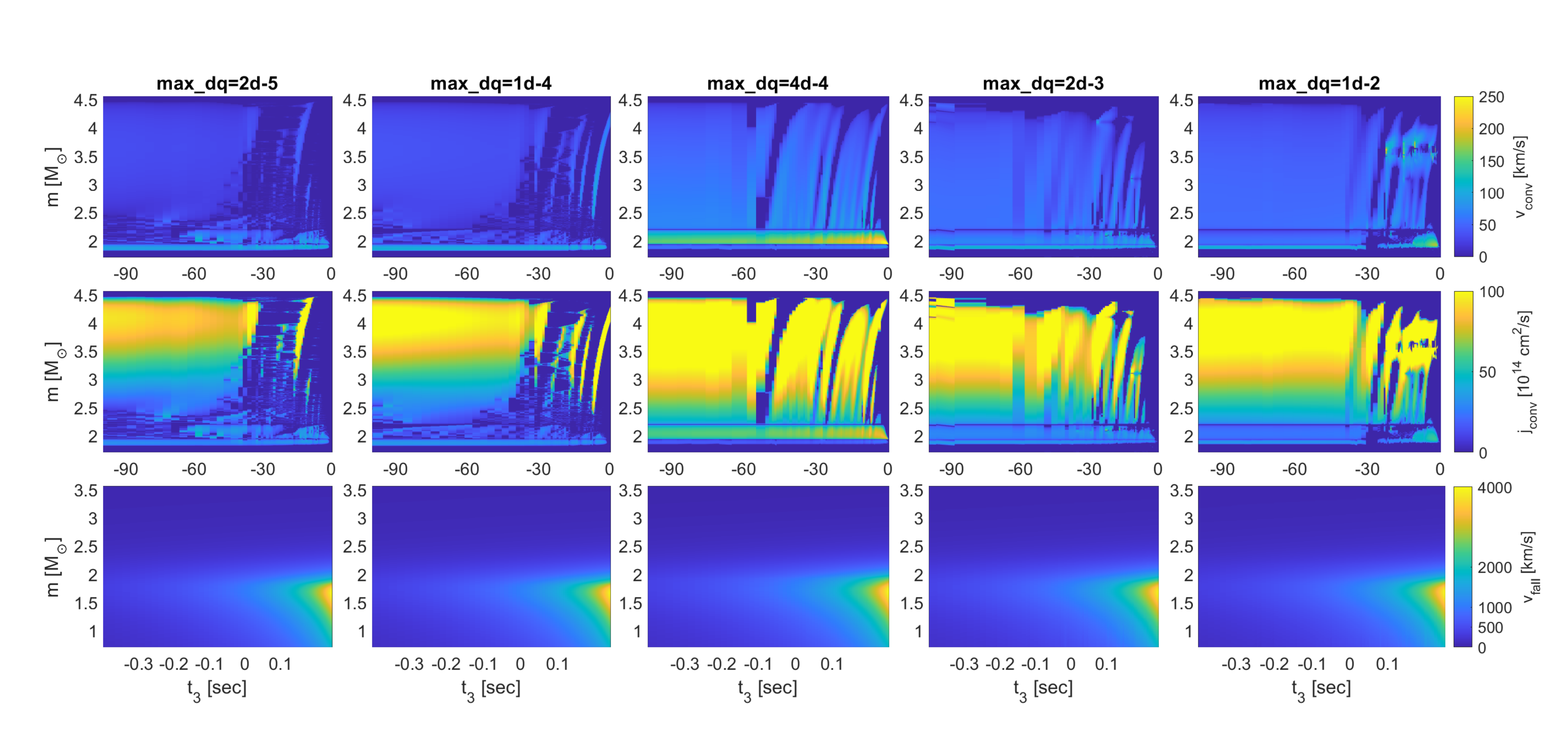}
\caption{The convective profile, i.e., the convective velocity $v_{conv}$ and its specific angular momentum fluctuations $j_{conv}$, and the infall velocity $v_{fall} (m)$), from upper to lower row respectively, as a function of mass and time $t_3$ for five different cell resolutions $max\_dq$ values in a $M_{ZAMS}=22 M_\odot$ stellar model. 
The values of the different quantities are according to the respective color-bars on the right for each row. The upper two rows starts at $t_3=-100 \s$ and the lower row starts at $t_3=-0.4 \s$, where $t_3=0$ when the maximum infall velocity is $v_{\rm fall,m}=1000 \km \s{-1}$. All graphs end when $v_{\rm fall,m} = 4000 \km \s^{-1}$, at about $t_3=0.2 \s$. }
\label{fig:App1fig}
\end{center}
\end{figure*}

We here present the role of the maximum allowed numerical cell mass in our simulations. For that we follow the evolution of the convective zones towards core collapse for the stellar model with $M_{ZAMS}=22 M_\odot$, starting from the time when the iron core mass is $M^{Fe}_{core}=1.2 M_{\odot}$. We stop the re-meshing once at least one cell reaches a temperature of $T> 6.3 \times 10^9 \K$.
We follow the evolution of three quantities in the core as function of mass and time. We present these in Fig. \ref{fig:App1fig}, where the vertical axis is the mass coordinate in the core and the horizontal axis is the time $t_3$, i.e., measured relative to the time when the maximum collapsing velocity is $v_{\rm fall,m}=1000 \km \s^{-1}$. The five columns in Fig. \ref{fig:App1fig} present the results for five different resolutions that we mark by the \textsc{mesa} numerical parameter $max\_dq$, which is maximum allowed numerical cell mass relative to stellar mass.   
   
In the upper to lower rows we display the evolution of the convective velocity, the specific angular momentum fluctuations $j_{conv}= v_{conv} r$, and the infall velocity, respectively. 
In the upper two rows dark-blue regions have no convection. 

From these evolutionary panels comparing high (left columns) to low (right columns) resolution we learn that although the different (maximal) cell sizes alter the convective profiles in the outer core (mass coordinates of $m \ga 2.5 M_\odot$), the overall picture in behavior of the inner convective zone which is the focus of our study and its evolution towards collapse remains similar between different resolutions. We conclude that the resolution that we use in this study $max\_dq = 1d-4$ is adequate for our goals.

\section{Convective zones}
\label{appendix2}

\begin{figure*}
\begin{center}
 \includegraphics[trim=1cm 2.5cm 1cm 1.8cm,scale=0.8]{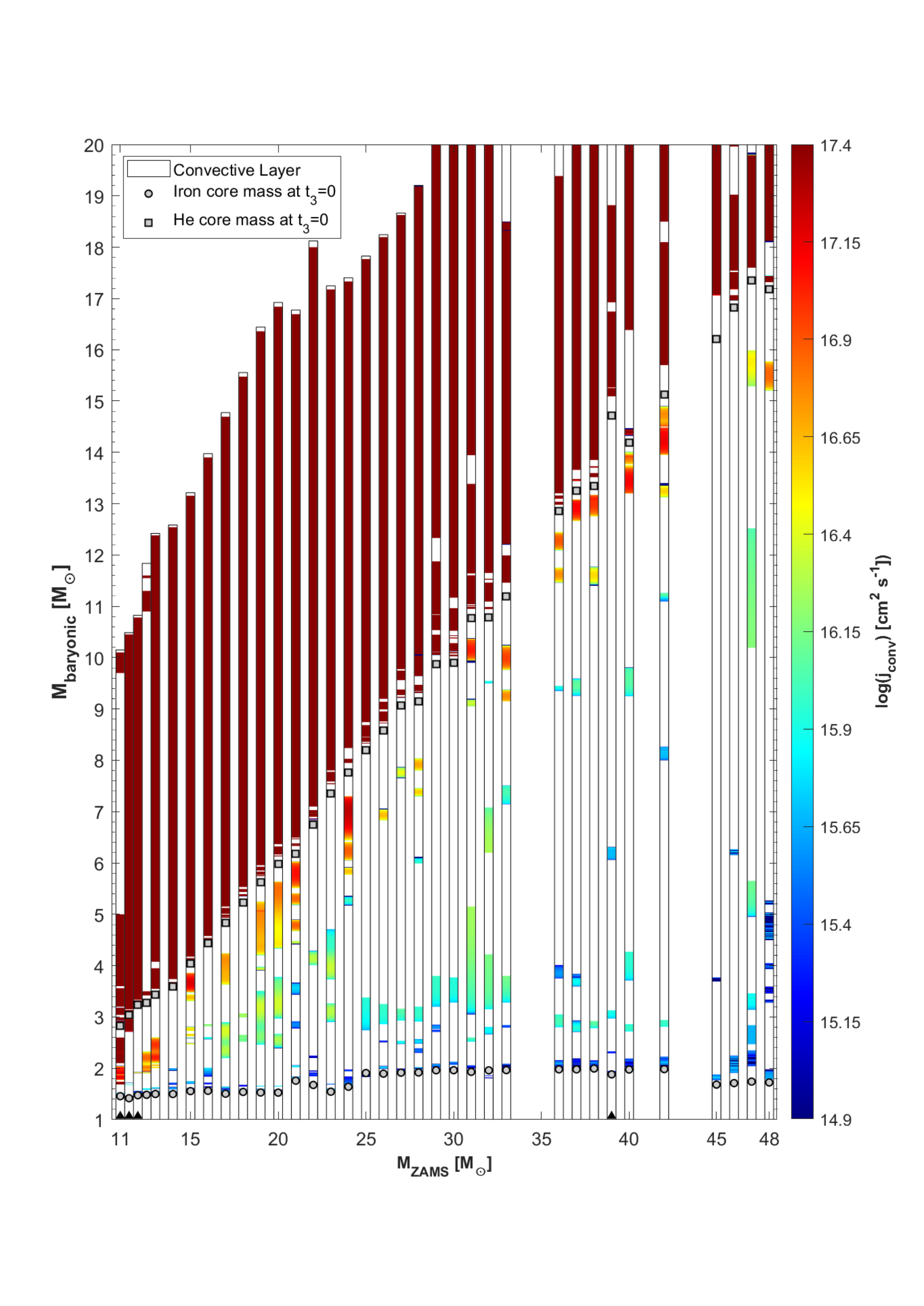}
\caption{The specific angular momentum fluctuations $j_{\rm conv}(m)$ in the different convective zones up to $M_{\rm baryonic}=m=20 M_\odot$. Colors represent the values of $j_{\rm conv}(m)$ according to the color-bar. All values of $j_{\rm conv}(m)> 10^{17.4} \cm^2 \s^{-1}$ have the color of $j_{\rm conv}(m) = 10^{17.4} \cm^2 \s^{-1}$.
Squares denote the helium core edge and circles the iron core edge. }
\label{fig:App2fig}
\end{center}
\end{figure*}

In Fig \ref{fig:App2fig} we present the full convective profiles of the $M_{\rm{ZAMS}}=11-48$ stellar models up to the $m=20 M_\odot$ (which is the baryonic mass). We display the values of $j_{\rm conv}(m)=v_{\rm conv}(m) r$ at each radius (unlike in Figs. \ref{fig:RemMass} and \ref{fig:RemMass2} where we averaged this value to obtain  $\bar{j}_{\rm conv}=\overline{v_{\rm conv} r}$ in each convective zone). Values of $j_{\rm conv}(m)$ are according to the color-bar.  

\end{document}